\documentclass{svjour3}                     % onecolumn (standard format)
\smartqed  % flush right qed marks, e.g. at end of proof
\usepackage{graphicx}
\journalname{myjournal}
\begin{document}

\title{Note on ``Continuous matter creation and the acceleration of the universe:
the growth of density fluctuations"}
%\subtitle{Do you have a subtitle?\\ If so, write it here}

\titlerunning{Note on growth of density fluctuations}        % if too long for running head

\author{J. A. S. Lima, J. F. Jesus and F. A. Oliveira}

%\authorrunning{Short form of author list} % if too long for running head

\institute{Instituto de Astronomia, Geof\'isica e Ci\^encias Atmosf\'ericas, Universidade de S\~ao Paulo\\
R. do Mat\~ao, 1226 - S\~ao Paulo, Brazil\\
\email{limajas@astro.iag.usp.br, jfernando@astro.iag.usp.br, foliveira@astro.iag.usp.br}%  \\
}

\date{Received: date / Accepted: date}
% The correct dates will be entered by the editor

\maketitle

\begin{abstract}

Recently, de Roany \& Pacheco \cite{deRoanyPacheco2010} performed a
Newtonian analysis on the evolution of perturbations for a class of
relativistic cosmological models with Creation of Cold Dark Matter
(CCDM) proposed by the present authors \cite{LJO}. In this note we
demonstrate that the basic equations adopted in their work  do not
recover the specific (unperturbed) CCDM model. Unlike
to what happens in the original CCDM cosmology, their basic
conclusions refer to a decelerating cosmological model in which there is no
transition from a decelerating to an accelerating regime as required
by SNe type Ia and complementary observations.

\end{abstract}
\keywords{cosmological fluctuations \and matter creation \and
accelerating universes}
% \PACS{PACS code1 \and PACS code2 \and more}
% \subclass{MSC code1 \and MSC code2 \and more}
%\end{abstract}
\vspace{0.5cm}

An alternative cosmological approach providing  a transition from an
early  decelerating to a late time accelerating expanding Universe
(as indicated by SN Ia data \cite{union}) and a reduction of the
so-called cosmic dark sector has been recently discussed in the
literature \cite{LJO,LimaEtAl2008,SteigmanEtAl2009}. The basic idea
is a simple one. The gravitationally-induced particle creation can
lead to an accelerating cold dark matter (CDM) dominated Universe
without requiring the presence of quintessence scalar fields or a
cosmological constant. This happens because the irreversible
creation of cold dark matter (CCDM) can thermodynamically be
described by a negative  pressure thereby resulting in a positive
acceleration as predicted by the Einstein  field equations
\cite{Pri89}. By neglecting the radiation and baryonic components, a
pure CCDM cosmology can be fully described by the following
relativistic equations \cite{LJO}
\begin{eqnarray}
\label{fried}
    8\pi G \rho &=& 3 \frac{\dot{a}^2}{a^2} + 3 \frac{k}{a^2},\\
\label{frw_p}
   8\pi G p_{c} &=&  -2 \frac{\ddot{a}}{a} - \frac{\dot{a}^2}{a^2} -\frac{k}{a^2},
\end{eqnarray}
where $\rho$ is the CDM density, $p_c$ is the creation pressure, and
an  overdot means time derivative. In the case of constant specific
entropy per particle (``adiabatic'' particle creation), the creation
pressure for a CDM component is given by \cite{Pri89}
\begin{equation}
\label{CP}
    p_{c} = -\frac{\rho\Gamma}{3H},
\end{equation}
where $H = {\dot {a}}/a$ is the Hubble parameter  and $\Gamma$ is
the creation rate of CDM particles.

Now, following standard lines, the above equations can be rewritten
in order to obtain the  ``continuity"  and acceleration equations:
\begin{eqnarray}
\label{ConsDM}
{\dot{\rho}} + 3H{{\rho}} &=& {{\rho}}\Gamma,\\
 \label{AccelCCDM}
\frac{\ddot{a}}{a}=-\frac{4\pi G}{3}(\rho+3p_c)&=&-\frac{4\pi
G\rho}{3}\left(1-\frac{\Gamma}{H}\right).
\end{eqnarray}
The above Eq. (\ref{AccelCCDM}) shows that creation of CDM
 particles ($\Gamma>0$) may provide a transition from a decelerating
to an accelerating stage. Note also that the  relativistic
description of a CCDM cosmology is fully determined once the
$\Gamma$ parameter is given.  In the specific CCDM cosmology
proposed by Lima et al. \cite{LJO} (from now on LJO model), the
creation rate is defined by:

\begin{equation}
\label{gamma} \frac{\Gamma}{H}=3 \alpha\frac{\rho_{c0}}{\rho},
\end{equation}
where $\rho_{c0}=3H_0^{2}/8 \pi G$ is the present value of the
critical density. Inserting the above expression into (\ref{CP}) one finds for the creation pressure, $P_c= -\alpha \rho_{co}$, which is negative and constant 
(termed -$\lambda$ in the notation of \cite{deRoanyPacheco2010}). At the level of background equations, it has  also been
proved that the cosmic history of such a CCDM cosmology is
indistinguishable from the standard $\Lambda$CDM model. As one may
check,   by combining Eqs. (\ref{CP}) and (\ref{gamma}) with  the
second Friedman equation (\ref{frw_p}), it is readily seen that the
evolution of the scale factor for LJO model reads:

\begin{equation}
\label{evolR} 2a{\ddot a}+ {\dot{a}}^2 + k - 3\alpha{H_0}^{2} a^{2}
= 0,
\end{equation}
which should be compared to:
\begin{equation}
\label{evolRLCDM} 2a{\ddot a}+ {\dot{a}}^2 + k - {\Lambda}a^{2}  =
0,
\end{equation}
provided by the $\Lambda$CDM model. The above equations imply that
the LJO model has the same cosmic dynamics of a $\Lambda$CDM
Universe when we identify the creation parameter by the expression
$\alpha = {\Lambda}/3{H_0}^{2} \equiv \Omega_{\Lambda}$.  In
particular, the LJO model  predicts the same $\Lambda$CDM transition
from a decelerating to an accelerating regime with two basic
advantages: (i) the cosmological constant problem is avoided, and
(ii) the dark sector is reduced to a simple component (cold dark
matter). The price to pay is the lack of a proper quantum field
theoretical approach for irreversible creation of cold dark matter.
Until the present, such a mechanism has been consistently justified
only in terms of nonequilibrium relativistic thermodynamics
\cite{Pri89} and kinetic theory \cite{Kinet}.

On the other hand, despite that both models ($\Lambda$CDM and LJO)
may share an identical Hubble expansion history, the same could not happen at
higher orders on the theory of small density fluctuations.
Therefore, it would be of interest to analyze the evolution of small
density perturbations by taking into account the matter creation
process. This is what de Roany and Pacheco proposed to analyze in
their paper. By adopting the Poisson and Euler equations together a
modified continuity equation (in order to include CDM creation) they
performed a Newtonian analysis of the small density perturbations in
the framework of the LJO  model. However, as we shall see below, the
basic equations assumed in their analysis fail to recover the
accelerating LJO model. In other words, even  considering that the
work is correct from a mathematical viewpoint, all the criticism in
their paper refer to a decelerating cosmology and not for the
scenario proposed by the authors. This is the basic result derived
in the present note. At this point, we stress that  the original
notation of \cite{LJO} will be adopted. In the notation of
\cite{deRoanyPacheco2010}, $\alpha$ is represented by $\Omega_v$.

%\section{Roany \& Pacheco equations}
The Newtonian analysis\footnote{It should be recalled that a nonrelativistic approach works when the scale of the perturbations is much less than the
Hubble radius and the velocity of peculiar motions are small in comparison to the Hubble flow \cite{Peebles}.} of de Roany and Pacheco
\cite{deRoanyPacheco2010} starts with a modified continuity equation
(in order to include creation of CDM) combined with Euler and
Poisson equations:
\begin{eqnarray}
\label{pconserv}
\frac{\partial \rho}{\partial t} + \vec{\nabla} \cdot (\rho \vec{U})&=& \rho \Gamma,\\
\label{peuler} \frac{\partial \vec{U}}{\partial t} + (\vec{U} \cdot
\vec{\nabla})\vec{U} &=& -\vec{\nabla} \phi,\\
\label{ppoten}
\nabla^2 \phi &=& 4 \pi G \rho(t),
\end{eqnarray}
where $\rho(t)$ is the CDM density, $\vec{U}$ is the velocity field
of the fluid and  $\Gamma$ is the creation rate  as defined in the
LJO model (see Eq. (\ref{gamma})).

Let us try to see if the basic relativistic LJO equations are
recovered from the above set of equations. To begin with, we recall
that isotropy and spatial homogeneity of the unperturbed model
implies that $\vec U = H{\vec r}$ (Hubble's law). In addition, since
$\vec{\nabla} \cdot \vec{U}=\vec{\nabla} \cdot H\vec{r}=3H$, one may
rewrite Eq. (\ref{pconserv})  as:
\begin{equation}
\label{rhodm} \dot{\rho}+3H\rho = \rho  \Gamma
\end{equation}
which is exactly equation (\ref{ConsDM}) of LJO. In addition, by
inserting the expression of $\Gamma$ as given by (\ref{gamma}), the
above equation can directly be integrated. Apart a slightly
different notation, the solution given by de Roany and Pacheco reads
(see Eq. (13) in \cite{deRoanyPacheco2010})
\begin{equation}
\label{rhoa}
 \rho=(\rho_0-\alpha\rho_{c0})a^{-3}+\alpha\rho_{c0},
\end{equation}
which also coincides with the solution of LJO  (see Eq. (9) in
\cite{LJO}). Let us now consider the Euler equation (\ref{peuler})
which can be rewritten as:
\begin{equation}
\label{eqeuler2} \dot{H}\vec{r} + H r \frac{\partial}{\partial r}(H
\vec{r})= \left( \dot{H} + H^2 \right) \vec{r} = -\vec{\nabla} \phi.
\end{equation}
where the right hand side is defined by the Poisson equation
(\ref{ppoten}). As one may check, since the density depends only on the time, an integration of the Poisson equation (\ref{ppoten}) yields 
\begin{equation}
\label{PoissonInt} \vec{\nabla} \phi = 4 \pi G \rho
\frac{\vec{r}}{3},
\end{equation}
where the integration constant was chosen in such a way that the
gravitational field is null in the center of the distribution.
Finally, by inserting the identity $\dot{H} = {\ddot a}/ {a} - H^2$ in
(\ref{eqeuler2}) and using (\ref{PoissonInt}) we derive the
acceleration formula:
\begin{equation}
\label{pfriedmann} \frac{\ddot{a} }{a} = -\frac{4 \pi G}{3} \rho ,
\end{equation}
which is fully different from Eq. (\ref{AccelCCDM}) of LJO model.
This result means that the basic equations adopted by de Roany and
Pacheco describe a decelerating model.  In particular, their
description does not permit a transition from a decelerating to an
accelerating Universe as required by SNe Ia observations. Naturally,
for all practical purposes the note finished here. However, for
completeness, it is interesting to know which is the Friedmann
equation for the energy density in the kind of model discussed by de
Roany and Pacheco.

By inserting Eq. (\ref{rhoa}) into (\ref{pfriedmann}) and
multiplying the result by $\dot a$, a simple integration yields
%\begin{equation}
%\label{}
%d{(\dot{a}^ 2) }= - \frac{8 \pi G}{3} a \rho da ,
%\end{equation}
\begin{equation}
\label{acc} {\dot{a}^ 2 +  k} =  \frac{8 \pi G}{3} \left[ (\rho_0 -
\alpha \rho_{c0}) a^ {-1} - \frac{1}{2} \alpha \rho_{c0} a^ 2
\right],
\end{equation}
where k is an arbitrary integration constant. Finally, dividing both
sides by $a^{2}$ and using again Eq. (\ref{rhoa}) together
(\ref{CP}) and (\ref{gamma}), it follows that
\begin{equation}
\label{FR1} {8 \pi G} \left[ \rho + \frac{3}{2}p_c \right]= 3
\frac{\dot{a}^2}{a^{2}} + 3\frac{k}{a^2},
\end{equation}
which should be compared with the Friedman equation (\ref{fried}).
In the above equation there is a spurious extra contribution, $12\pi
Gp_c$. Its presence confirms again that the basic cosmological equations describing
the unperturbed  LJO model cannot be recovered with basis on the
adopted nonrelativistic approach. As recently discussed by Basilakos and Lima \cite{BL10}, a consistent 
quasi-Newtonian treatment with pressure also requires a modification of the Poisson equation (see also \cite{LZB97}). 

%This contribution is negative, giving rise to the deceleration.
%The equation of motion can be written as:
%\begin{equation}
%\label{pmov}
%2 \ddot{a}a + \dot{a}^2 + (4 \pi G \alpha \rho_{c0}) a^ 2 + k = 0 .
%\end{equation}

In conclusion, we have shown that the Newtonian analysis performed
by de Roany and Pacheco \cite{deRoanyPacheco2010} fails when trying to recover the background
(zero order) relativistic model proposed in Ref. \cite{LJO}. All
their results describing linear perturbations seems to be mathematically
correct, however, they refer to a decelerating cold dark matter
Universe with a modified continuity equation.  It is also clear that by using different background equations 
will result on a different evolution for the linear density contrast since the corrections will generate new terms  on the first 
order perturbation equations. A more rigorous ``Newtonian formulation" for the relativistic LJO models, the corresponding evolution of small
density fluctuations and other physical consequences is being developed and will be
published elsewhere \cite{JOBL}.
\\
\\
{\bf Acknowledgments:}
The authors would like to thank Spyros Basilakos for helpful discussions  and A.~de Roany and
 J.~A.~de Freitas Pacheco for useful correspondence. JFJ is supported by Fapesp, FAO is supported by CAPES (Brazilian Research Agencies) and JASL is partially supported by CNPq and FAPESP under grants 304792/2003-9 and 04/13668-0, respectively.

\end{document}